\documentclass[preprint,prd, nofootinbib]{revtex4-1}
\usepackage{color}
\usepackage{graphicx}
\usepackage{epsfig}
\usepackage{subfig}
\usepackage{lipsum}
\usepackage{ctable}
\usepackage{hyperref}
\usepackage{physics}
\usepackage{graphicx}
\usepackage{dcolumn}
\usepackage{bm}

\begin{document}

	\title{Black Hole Solutions with Scalar Fields and Anisotropic Matter in Non-Minimal $Y(R)F^2$ Gravity}

	\author{ \"{O}zcan SERT}
	\email{osert@pau.edu.tr}
	\affiliation{Department of Physics, Faculty of  Sciences, Pamukkale	University,  20070   Denizli,
       T\"{u}rkiye  }
	

	\date{\today}

	\begin{abstract}

\noindent

We investigate exact  spherically symmetric and static black hole solutions in a non-minimally coupled Einstein--Maxwell theory of the $Y(R)F^2$ type, extended by a minimally coupled real scalar field and an anisotropic matter distribution. The Maxwell field is coupled to the spacetime curvature through an arbitrary function of the Ricci scalar, while the scalar field is described by a kinetic term and a scalar potential. We obtain exact solutions for power-law and logarithmic forms of the non-minimal coupling function and analyze the corresponding spacetime geometries. In particular, we identify a special $\beta=-1/3$ case for which the general power-law solution becomes singular and derive a new logarithmic metric by solving the field equations independently. The resulting configurations are asymptotically flat and  exhibit modified gravitational properties associated with the interplay between the scalar field, anisotropic matter, and non-minimal electromagnetic coupling. We further show that the magnetic solutions admit electrically charged counterparts through an electromagnetic duality transformation, under which the metric and scalar sector remain invariant while the non-minimal coupling function is inverted. The obtained solutions provide a useful framework for investigating black hole horizons, thermodynamic properties, geodesic motion, and astrophysical phenomena such as galactic rotation curves in curvature-dependent electromagnetic theories.

	\end{abstract}
	
	\pacs{Valid PACS appear here}
	\maketitle
 
 \section{Introduction}

The dynamics of galaxies and compact astrophysical objects provide important
laboratories for testing the gravitational interaction beyond the regime of the
Solar System. The rotation curves of spiral galaxies provide one of the most persistent indications of a possible discrepancy between the observed distribution of matter and the gravitational dynamics predicted by General Relativity. Observations show that the circular velocities of stars and gas in the
outer regions of many spiral galaxies remain approximately constant over a
substantial range of galactocentric distances, whereas the contribution of the
visible baryonic matter is expected to decrease approximately according to a
Keplerian behavior once the bulk of the luminous matter has been enclosed.
This discrepancy has traditionally been interpreted as evidence for extended
dark-matter halos surrounding galaxies
\cite{Rubin1980,Bosma1981,Begeman1991,Trimble1987}.

The dark-matter paradigm provides a remarkably successful framework for
describing a wide range of cosmological and astrophysical observations.
Nevertheless, the fundamental nature of dark matter remains elusive, and this
has motivated extensive investigations of alternative gravitational scenarios.
In particular, modified theories of gravity have been explored as possible
explanations of gravitational phenomena that are conventionally attributed to
dark matter. Among these approaches are  $f(R)$
gravity, scalar-tensor theories, symmetric teleparallel gravity, and other extensions of General
Relativity involving additional gravitational degrees of freedom
\cite{Clifton2012,Brax2008,SotiriouFaraoni2010,AdakSert2005,AdakKalaySert2006}. These theories provide different ways of modifying
the relation between the gravitational field and the distribution of matter,
and some of them can reproduce the approximately flat behavior of galactic
rotation curves in suitable regimes.

Another possibility is to describe the additional gravitational contribution
through effective fields or fluids. Scalar-field configurations, in
particular, have been extensively investigated as possible sources of
galactic halos and as alternatives to conventional particle dark matter
\cite{Matos2000,Arbey2003,Matos2001}. Quintessence-inspired configurations
have also been considered in connection with the asymptotic behavior of
galactic rotation curves. In such descriptions, the effective equation-of-state
parameter of the additional component can play an important role in determining
the large-distance behavior of the gravitational field
\cite{Kiselev2003}. These studies demonstrate that additional gravitational
degrees of freedom and non-standard matter sources can significantly modify
the geometry of spacetime and lead to phenomenological effects that are absent
in the minimally coupled Einstein theory.

A particularly interesting extension of General Relativity is obtained by
allowing matter fields to interact non-minimally with the curvature of
spacetime. Such curvature--matter couplings can modify the effective
gravitational source while retaining the geometric framework of the theory \cite{Bertolami2007}.
In particular, non-minimal interactions between electromagnetic fields and
curvature have attracted considerable attention \cite{Prasanna1971,Horndeski1976,Drummond1980,Dereli19901,
Buchdahl1979,Muller-Hoissen1988,Turner1988,Lambiase2004,
Mazzitelli1995,Campanelli2008,Lambiase2008,Bamba20082,Bamba20081,
Bamba2007,Dereli20111,DereliSert2011MPLA}. In the minimally coupled
Einstein--Maxwell theory, the electromagnetic field interacts with gravity
through the standard Maxwell invariant. In contrast, a curvature-dependent
coupling allows the electromagnetic contribution to the gravitational field
equations to depend explicitly on the spacetime curvature. Consequently, the
resulting gravitational configurations can exhibit properties that differ
from those of the standard Reissner--Nordstr\"om geometry.

A particularly interesting class of theories is described by the non-minimal
$Y(R)F^2$ coupling, where $F$ is the electromagnetic field strength and
$Y(R)$ is a function of the Ricci curvature scalar $R$. In these models, the
electromagnetic sector is directly sensitive to the spacetime curvature \cite{Dereli2007,Dereli20072},
leading to modified gravitational and electromagnetic field equations.
Different choices of the function $Y(R)$ have been employed to construct
cosmological, stellar, and black-hole solutions
\cite{Sert13MPLA,Sert12Plus,Sert2024Wormhole,
Dereli20112,Dereli20113,Sert2016,Sert2017,Sert2018,Sert20182,
AADS,SertAdak2019}.

In particular, the non-minimal coupling between the electromagnetic field and
the curvature has been used to obtain exact static and spherically symmetric
solutions. Power-law and logarithmic forms of the curvature-dependent
coupling are of particular interest because they generate non-trivial
corrections to the corresponding Einstein--Maxwell geometries \cite{Dereli20111,DereliSert2011MPLA}. These
constructions provide a useful framework for studying how the functional form
of the electromagnetic--curvature interaction affects the resulting spacetime
geometry.

Scalar fields provide another important mechanism for obtaining
self-gravitating configurations beyond the standard Einstein--Maxwell
system. They arise naturally in various gravitational and cosmological
models and can support non-trivial matter distributions. In particular,
scalar-field configurations have been extensively considered as effective
descriptions of galactic dark-matter halos and other self-gravitating systems
\cite{Matos2000,Li2012,Xu2019,BarbosaSantos2026,Ferreira1997,Ferreira1998,Copeland1998,
Billyard2000}. The presence of a scalar field introduces an additional
gravitational degree of freedom and can modify the effective energy density,
pressure distributions, and spacetime geometry.

Motivated by these considerations, it is natural to investigate the interplay
between scalar fields and curvature-dependent electromagnetic interactions.
In the present work, we extend the non-minimal Einstein-Maxwell
framework of the $Y(R)F^2$ type by including a minimally coupled real scalar
field together with an anisotropic matter distribution. The electromagnetic
field interacts with the gravitational sector through an arbitrary function
$Y(R)$, whereas the scalar field is coupled minimally to gravity through its
kinetic term and a scalar potential. 

An additional feature of the solutions is their relation to electrically
charged configurations. The electromagnetic sector of the theory admits a
 duality transformation that maps the magnetic solutions into
electric ones while transforming the non-minimal coupling function into its
inverse. Importantly, the spacetime geometry and the scalar sector remain
unchanged under this transformation. Thus, the electric configurations
provide dual counterparts of the magnetic solutions, with the metric,
scalar field, scalar potential, and anisotropic matter variables preserved.

The exact solutions obtained in this work provide new examples of
asymptotically flat geometries supported by the combined effects of a scalar
field, a non-minimally coupled electromagnetic field, and an anisotropic
matter distribution. These configurations may provide useful backgrounds
for studying horizon structure, thermodynamic properties, geodesic motion,
quasinormal modes, gravitational lensing, and stability in
non-minimally coupled gravity models. Furthermore, the presence of scalar
fields and anisotropic matter distributions makes these solutions potentially
relevant to astrophysical applications involving effective dark-matter
descriptions.

The paper is structured as follows. In Sec.~\ref{model}, we introduce the
coupled gravity model and derive the corresponding field equations. In the following sections, we consider a static, spherically symmetric spacetime in the presence of a magnetic field and derive exact solutions corresponding to power-law and logarithmic forms of the non-minimal coupling function. The corresponding electric configurations
obtained through electromagnetic duality are then discussed. Finally, we
summarize our main results and comment on possible future applications of the obtained 
 solutions.

\section{The Coupled Gravity Model} \label{model}

As an extension of our previous investigations
\cite{Sert13MPLA,Sert12Plus,Sert2024Wormhole,
Dereli20112,Dereli20113,Sert2016,Sert2017,Sert2018,Sert20182,
AADS,SertAdak2019}.
we introduce a real scalar field into the non-minimally coupled
$Y(R)F^2$ gravity model. In the present framework, the scalar field is
minimally coupled to gravity, whereas the electromagnetic field interacts
with the gravitational sector through the curvature-dependent coupling
function $Y(R)$.

The present model is motivated by the description of galactic halo
configurations within modified theories of gravity. In particular,
self-gravitating anisotropic matter distributions supported by scalar fields
have been extensively employed as effective models of galactic dark matter
halos, where the scalar field provides an additional long-range gravitational
degree of freedom capable of modifying the halo dynamics
\cite{Matos2000,Li2012,Xu2019,BarbosaSantos2026,Ferreira1997,Ferreira1998,Copeland1998,Billyard2000}.
Rather than describing the interior of compact stellar objects, we seek exact
static and spherically symmetric solutions representing effective anisotropic
matter distributions extending throughout spacetime. Such configurations may
provide useful toy models for galactic halos in the presence of non-minimal
curvature--electromagnetic couplings.

The corresponding Lagrangian 4-form is
\begin{eqnarray}\label{Model0}
L=
\frac{1}{2\kappa^2}R*1
-Y(R)F\wedge *F
-c_1 d\phi\wedge *d\phi
-V(\phi)*1
+L_m
+\lambda_a\wedge T^a .
\end{eqnarray}

Here, $Y(R)$ is an arbitrary function of the Ricci scalar $R$, $F$
denotes the electromagnetic field strength 2-form, $c_1$ is the
coupling constant, $\phi$ represents a scalar
field, and  $V(\phi)$ is a scalar potential. The term $L_m$
denotes the matter Lagrangian describing an anisotropic matter
distribution. The torsion-free condition, $T^a=0$, is enforced by the Lagrange multiplier 2-form $\lambda_a$, thereby restricting the connection to the torsion-free Levi--Civita connection and recovering the standard Riemannian geometric structure.
The corresponding energy--momentum 3-form is obtained by varying $L_m$ with respect to the coframe $e^a$,

\begin{eqnarray}
\tau_a^{m}
=
(\rho+p_t)u_a*u
+p_t*e_a
+(p_r-p_t)v_a*v .
\end{eqnarray}

Here, $\rho$ represents the energy density, while $p_r$ and $p_t$ denote the radial and tangential pressures, respectively. The timelike 1-form
$u=\delta^0_a e^a$ represents the four-velocity of a comoving observer, and
$v=\delta^1_a e^a$ is the unit spacelike 1-form in the radial direction. Throughout this work, the matter distribution is assumed to be static and anisotropic, and the observer is taken to be inertial and comoving with the fluid.

Applying the variational principle to the Lagrangian 4-form (\ref{Model0}) and varying it with respect to the coframe \(e^{a}\), one arrives at the gravitational field equations:
 \begin{align}
\label{gfe1}
- \frac{1}{2 \kappa^2}  R^{bc}
\wedge *e_{abc}  =   Y\tau_a(F)    -\frac{k}{\kappa^2}*R_a 
+      \tau_a^{mat} +     \tau_a(\phi)
 \end{align}
 under the following constraint
  \begin{eqnarray}\label{cond0}
 \frac{dY}{dR} F_{mn}F^{mn} = -\frac{ k}{ \kappa^2} \;. 
\end{eqnarray} 
Here $R_a=R_{a,b}e^b={R_{ac,b}}^ce^b$ is the Ricci 1-form, and
the energy-momentum tensor of the electromagnetic field and the scalar field are  \begin{eqnarray}
\tau_a(F)    &=& F_a \wedge *F - F \wedge \iota_a *F \;,\\
\tau_a(\phi) &=& c_1(\iota_a d\phi\wedge*d\phi  + d\phi\wedge \iota_a *d\phi)  - V(\phi)*e_a \;. 
\end{eqnarray}
Throughout this study, we also denote $Y_R= \frac{dY(x)}{dR(x)}$.
The constraint (\ref{cond0}) plays a central role: it simplifies the model and allows the construction of non-trivial exact solutions in non-minimally coupled theories \cite{Sert2016,Sert2017,Sert2018}. For $k=0$, the coupling function $Y$ becomes constant and the theory reduces to the standard Einstein–Maxwell system with minimal coupling. This constraint also simplifies the gravitational sector and ensures ghost-free dynamics, as the resulting field equations contain derivatives of at most second order. For further discussion of its consistency and implications, see Ref.~\cite{Sert2017}.

An equivalent formulation is obtained by imposing (\ref{cond0}) at the level of the action via a Lagrange multiplier added to (\ref{model1}). Variation of the resulting extended action reproduces the field equations in the form (\ref{gfe1}). The gravitational field equation (\ref{gfe1}) can then be written equivalently as follows:
\begin{eqnarray}\label{gfe3}
    G_a=\kappa^2\tau_a^{eff}
\end{eqnarray}
where  
$G_a = - \frac{1}{2 \kappa^2}  R^{bc}   \wedge *e_{abc} = *R_a-\frac{1}{2}R*e_a$ is the Einstein tensor
and $ \tau_a ^{eff}$ is   the effective energy-momentum tensor  3-form given by  
\begin{eqnarray}\label{tauabeff}
   \tau_a ^{eff}&=&    Y\tau_a(F)    -\frac{k}{\kappa^2}*R_a 
+      \tau_a^{mat} +     \tau_a(\phi)   = \tau_{ab}^{eff}*e^b
\end{eqnarray}
Here we  define that $\tau_{00}^{eff}= \rho^{eff}$ is the effective energy density,  $\tau_{11}^{eff}= P_{r}^{eff}$ is the effective radial pressure,  ${\tau_{22}}^{eff}= \tau_{33}^{eff}= P_{t}^{eff}$ is the effective tangential  pressure. The Bianchi identity, $DG_a=0$, guarantees the conservation of the effective energy-momentum tensor, which satisfies $D\tau_a^{\mathrm{eff}}=0$.

By taking  scalar field $\phi$ and electromagnetic potential  $A$ variation of the Lagrangian (\ref{model1}), we obtain 
\begin{eqnarray}
2c_1d*d\phi  -\frac{dV}{d\phi}*1=0 \label{phi1}\\
d(*Y F) = 0, \hskip 1 cm dF=0 
\label{Maxwell1}
\end{eqnarray} 
respectively.
Furthermore, taking the trace of the gravitational field equation (\ref{gfe1}) by wedging it with $e^a$, we obtain
 \begin{eqnarray} \label{trace}
 \frac{1-k}{\kappa^2} R*1 =2c_1d\phi\wedge *d\phi + (\rho  - p_r -2p_t + 4V)*1 \;.
 \end{eqnarray}

\section{SPHERICALLY SYMMETRIC, STATIC SOLUTIONS}

We take the following  static, spherically symmetric metric ansatz to obtain asymptotically flat solutions to the non-minimal model with scalar field. 
\begin{eqnarray}\label{metric}
ds^2 & =& -f(r)dt^2  + \frac{dr^2}{f(r)} + r^2d\theta^2 +r^2\sin^2\theta d \phi^2 
\end{eqnarray}
Consistent with this geometry, the electromagnetic tensor  has only the electric and magnetic field components that depend on the radial coordinate.   
\begin{eqnarray}
\label{Maxwellanzats}
 F &= & E(r) e^1\wedge e^0 +B(r)e^2\wedge e^3\; .
\end{eqnarray}

In particular, solutions with only electric component $E(r)$ or only magnetic component $B(r)$ can be obtained by taking $B=0$ or $E=0$ respectively, depending on the source.
With this non-minimal model, we can define the constitutive tensor $G$ as the Maxwell tensor in the gravitational medium \cite{Dereli2007,Dereli20072};
\begin{eqnarray}
G=YF = YE e^1\wedge e^0 + YBe^2\wedge e^3 \;.
\end{eqnarray}
Here   $D=YE$  corresponds to the displacement field  and $H=YB$ the magnetic field  in the specific gravitational medium. 
The modified electromagnetic field equations (\ref{Maxwell1}) for these metric and vector field ansatzes become
\begin{eqnarray}\label{Maxwell2}
 D(r)=   Y(r)E(r) = \frac{q_e}{r^2}\;, \hskip 1.5 cm B(r)=\frac{q_m}{r^2}\;.
\end{eqnarray}
In these equations, the integration constants $q_e$ and $q_m$ are identified with the electric and magnetic monopole charges of the source, respectively.
These metric and vector field ansatzes yield the following system of  differential equations for the  field equation (\ref{gfe3})  of the non-minimal model.
 \begin{eqnarray}
   -\frac{f'}{r} -\frac{f}{r^2} +\frac{1}{r^2} &=&\rho^{eff} 
   \;, \label{d1} 
\\
 \frac{f'}{r} +\frac{f}{r^2} -\frac{1}{r^2} &=& P_r^{eff}
   \;,  \label{d2}
\\
  \frac{f''}{2}   + \frac{f'}{r}   &=&  P_t^{eff}   \;, \label{d3}
\end{eqnarray}
subject to the constraint that
\begin{eqnarray}
    \frac{dY}{dR}(E^2-B^2) = \frac{k}{2\kappa^2}\label{constraint}\;.
\end{eqnarray}
In these differential equations (\ref{d1},\ref{d2},\ref{d3}) effective energy density and pressures are 
\begin{eqnarray}
  \rho^{eff} &=& \kappa^2(c_1f\phi'^2 + V)- k( \frac{f''}{2} + \frac{f'}{r}  ) +  \kappa^2 Y( E^2+B^2)     + \kappa^2\rho \;, \\
    P_r^{eff}&=& \kappa^2(c_1f\phi'^2-V)+ k( \frac{f''}{2} + \frac{f'}{r}  )   
- \kappa^2 Y( E^2+B^2)     +  \kappa^2 p_r \;,
    \\
     P_t^{eff}  &=& 
     -\kappa^2(c_1f\phi'^2+V)+ k( \frac{f'}{r} + \frac{f}{r^2} -\frac{1}{r^2} )  +\kappa^2 Y( E^2+B^2)     +  \kappa^2 p_t \;.
\end{eqnarray}
Firstly, by  adding equations (\ref{d1}) and (\ref{d2}), and subsequently combining equations (\ref{d2}) and (\ref{d3}) side by side, we obtain the following results:
\begin{eqnarray}
\label{add1}
\rho^{eff} +  P_r^{eff}= 2c_1f\phi'^2 +\rho+p_r=0 
\end{eqnarray}
\begin{eqnarray}\label{add2}
    \frac{1-k}{\kappa^2}R=-2p_r-2p_t +4V
\end{eqnarray}
where  the Ricci curvature scalar $R$ is
\begin{eqnarray}
    R = -f'' -\frac{4f'}{r} -\frac{2(f-1)}{r^2} 
\end{eqnarray} for the spherically symmetric metric.
We have also
\begin{eqnarray}\label{trace2}
    \frac{1-k}{\kappa^2}R =2c_1\phi'^2f +\rho - p_r -2p_t+4V
\end{eqnarray}
from the trace equation (\ref{trace}). This result (\ref{trace2}) is also obtained  by adding these two equations (\ref{add1}) and (\ref{add2}).
Scalar field equation (\ref{phi1}) gives the following differential equation for the radial scalar field $\phi(r)$  and potential $V(r)$ 
\begin{equation}
2c_1\phi'(r)f'(r)
+\frac{4c_1f(r)}{r}\phi'(r)
+2c_1f(r)\phi''(r)
-\frac{V'(r)}{\phi'(r)}
=0.
\label{scalar-dif}
\end{equation}

\section{Magnetically Charged solutions}

Magnetically charged configurations are of particular interest in these models, as they provide a natural setting for investigating the effects of non-minimal electromagnetic-gravitational coupling on spacetime geometry and offer a convenient route to obtaining electrically charged solutions. Furthermore, they provide valuable insight into the behavior of nonlinear electrodynamics in strong gravitational fields, and have applications in early-universe cosmology, astrophysical compact objects, and holographic models within the AdS/CFT framework.
To proceed, we first restrict our investigation to the case of a purely magnetic configuration by setting $ q_e = 0$, which yields a vanishing electric field, $E = 0$. The magnetic field, consistent with spherical symmetry, is then given by
\begin{eqnarray}
    B = \frac{q_m}{r^2},
\end{eqnarray}

where $q_m$  denotes the magnetic monopole charge. Under this assumption, and within a static and spherically symmetric spacetime background, the field equations reduce to a simplified system that admits the following exact solution:
\begin{eqnarray}
    Y(r) &=& \frac{k}{2q_m^2\kappa^2}(f''r^4-2fr^2+2r^2) +C_1 \label{Y1r}\\
    V(r)&= & 2c_1\int \phi'[f\phi'' +\phi'(f'+2f/r)] dr   + C_2 \label{Vr}\\
    \rho(r) &=& -c_1f\phi'^2 - V +\frac{k}{\kappa^2}(\frac{f''}{2} +\frac{f'}{r} ) -\frac{q_m^2Y(r)}{r^4} -\frac{1}{\kappa^2}(\frac{f'}{r} +\frac{f-1}{r^2})\label{rhor}\\
    p_r(r) &=& -c_1f\phi'^2 + V - \frac{k}{\kappa^2}(\frac{f''}{2} +\frac{f'}{r} ) +\frac{q_m^2Y(r)}{r^4} 
    +\frac{1}{\kappa^2}(\frac{f'}{r} +\frac{f-1}{r^2})\label{prr}\\
        p_t(r) &=& c_1f\phi'^2 + V - \frac{k}{\kappa^2}(\frac{f'}{r} + \frac{(f-1)}{r^2}) -\frac{q_m^2Y(r)}{r^4} 
    +\frac{1}{\kappa^2}(\frac{f''}{2}+\frac{f'}{r} )\label{ptr}
\end{eqnarray}

\subsection{Power-Law Model with Magnetically Charged Solutions}

To obtain magnetically charged solutions in the present model, we first
specify the functional form of the non-minimal coupling function $Y(R)$
governing the interaction between the gravitational and electromagnetic
fields. We first consider the power-law coupling \cite{DereliSert2011MPLA}
\begin{equation}
Y(R)=1-R_0R^\beta,
\end{equation}
where $R_0$ and $\beta$ are real coupling parameters. This form represents
the lowest-order nonlinear correction to the Einstein--Maxwell theory while
remaining analytically tractable. It also enables us to investigate how the
electromagnetic sector is modified by the spacetime curvature and how these
modifications affect magnetically charged black-hole solutions.

The corresponding Lagrangian 4-form is chosen as
\begin{equation}
\label{model1}
L=
\frac{1}{2\kappa^2}R*1
-\left(1-R_0R^\beta\right)F\wedge *F
-c_1d\phi\wedge *d\phi
-V(\phi)*1
+L_m
+\lambda_a\wedge T^a .
\end{equation}
In the limit $R_0\rightarrow0$, the model reduces to the standard
Einstein--Maxwell theory minimally coupled to a scalar field. For
nonvanishing $R_0$, the curvature-dependent interaction modifies both the
electromagnetic field equations and the effective gravitational dynamics,
leading to deviations from General Relativity in strong-field regimes.
Such curvature-dependent non-minimal couplings have been extensively
investigated as a possible mechanism for the generation and amplification
of primordial magnetic fields in the early universe
\cite{Turner1988,Lambiase2004}. Although the simplest interaction
$RF\wedge *F$ leads to magnetic field amplification, the resulting field
strength is generally insufficient to account for present observations.
More general curvature-dependent couplings, including
$R^\alpha F\wedge *F$ and $R^{-\alpha}F\wedge *F$, have been shown to
significantly enhance primordial magnetic fields and provide a richer
phenomenology in both cosmology and strong gravitational fields
\cite{Mazzitelli1995,Campanelli2008,Lambiase2008,Bamba20081,Bamba20082}.
Moreover, the general conditions for obtaining sufficiently large
primordial magnetic fields have been formulated in terms of an arbitrary
non-minimal function $Y(R)$ \cite{Bamba2007}.
In the following, we investigate static, spherically symmetric magnetically
charged solutions arising from this model.

Solving the differential equation (\ref{Y1r}) for the model (\ref{model1})
leads to the following explicit form of the metric function:
\begin{eqnarray}\label{metric1}
f(r) &=&
1-\frac{2M}{r}+\frac{q_m^2}{r^2}-a_1r^\alpha .
\end{eqnarray}
Here $M$ is the integration constant, while $a_1$ and $\alpha$ are given by
\begin{eqnarray}
a_1 &=&
\left(
\frac{k}{2q_m^2\kappa^2\beta R_0}
\right)^{\frac{1}{\beta-1}}
\frac{(\beta-1)^2}{4\beta(3\beta+1)},
\label{a1beta}
\\
\alpha &=&
\frac{2\beta+2}{\beta-1}.
\end{eqnarray}
The above solution is valid for
$\beta\neq0,-1/3,1$, while $\alpha<0$ ensures asymptotic flatness.
Thus, the asymptotically flat branch is restricted to
$-1<\beta<1$, with $\beta\neq0,-1/3$. Since the coefficient $a_1$
becomes singular at $\beta=-1/3$, this special case must be treated
separately and is discussed in the following subsection. The case
$\beta=-1/3$ is special because it corresponds to $\alpha=-1$, for which
the Ricci scalar
\begin{eqnarray}\label{R1}
R(r)=a_1(\alpha+2)(\alpha+1)r^{\alpha-2}.
\end{eqnarray}
vanishes identically. Consequently, the power-law matching becomes
degenerate, and the $\beta=-1/3$ branch has to be constructed separately.
For the nondegenerate branch, Eq.~(\ref{R1}) can be inverted to express
the radial coordinate in terms of the Ricci scalar:
\begin{eqnarray}
r(R)=
\left[
\frac{R}
{a_1(\alpha+2)(\alpha+1)}
\right]^{\frac{1}{\alpha-2}}.
\end{eqnarray}
Matching the $r$-independent constant part of $Y(r)$ in Eq.~(\ref{Y1r})
with the corresponding constant term in
$Y(R)=1-R_0R^\beta$ fixes the integration constant as
\begin{equation}
C_1=1-\frac{2k}{\kappa^2},
\end{equation}
while the $r$-dependent parts of the two expressions are consistently
identified through the relation
\begin{equation}
\alpha+2=\beta(\alpha-2),
\end{equation}
reproducing the coefficient $a_1$ given in Eq.~(\ref{a1beta}).

To obtain solutions of the differential equations
(\ref{Vr},\ref{rhor},\ref{prr},\ref{ptr}) resulting from the field equations,
we adopt a suitable ansatz for the real scalar field motivated by
Lifshitz-like geometries
\cite{Cadoni20111,Cadoni20112,Charmousis2009,Goldstein2009,
Goldstein2010,Garfinkle1991}. In such configurations, the scalar field
can exhibit a logarithmic dependence on the radial coordinate, which we
take in the form
\begin{equation}
\phi(r)=\phi_0\ln\!\left(\frac{r}{r_0}\right).
\label{skalarfonk}
\end{equation}
Such logarithmic scalar profiles commonly arise in scale-invariant
Einstein--scalar systems and in a variety of scalar--tensor and
dilaton-like models. In the present construction, this ansatz is
particularly convenient because
\[
\phi'(r)=\frac{\phi_0}{r},
\]
so that the scalar kinetic contribution scales as $r^{-2}$ and remains
compatible with the power-law metric ansatz.

This scalar field is naturally compatible with power-law metric functions
and facilitates the construction of exact or semi-analytic solutions,
particularly in the presence of exponential scalar potentials or
non-minimal curvature couplings. Such structures arise frequently in
string-inspired models and effective gravitational actions, including
dilaton gravity and generalized scalar--tensor theories
\cite{Garfinkle1991,Gibbons1988,Damour1993,Polchinski1998,Sotiriou2012}.
Although logarithmic scalar fields do not asymptote to a constant value at
spatial infinity and therefore do not satisfy the standard assumptions
underlying no-hair arguments, they can provide consistent effective
descriptions over finite radial domains. Consequently, such configurations
have been employed in studies of scalar hair, anisotropic stress--energy
tensors, and large-scale gravitational phenomenology, including galactic
halo modeling and astrophysical modifications of General Relativity
\cite{Matos2000,Li2012,Xu2019,BarbosaSantos2026,Ferreira1997,Ferreira1998,
Copeland1998,Billyard2000}.

The scalar potential obtained from Eq.~(\ref{Vr}) is
\begin{eqnarray}
V(r)=
\frac{c_1\phi_0^2}{r^2}
\left(
-1+\frac{q_m^2}{2r^2}
-\frac{2a_1(\alpha+1)}{\alpha-2}r^\alpha
\right)
+C_2.
\end{eqnarray}
For the asymptotically flat branch, all $r$-dependent contributions in
$V(r)$ vanish as $r\rightarrow\infty$. We therefore set
\begin{equation}
C_2=0,
\end{equation}
so that the scalar potential also vanishes asymptotically.
Using the scalar field (\ref{skalarfonk}), the potential can be written
in terms of $\phi$ as
\begin{eqnarray}
V(\phi)
&=&
\frac{c_1\phi_0^2}{r_0^2}
e^{-2\phi/\phi_0}
\left[
-1
+\frac{q_m^2}{2r_0^2}
e^{-2\phi/\phi_0}
-\frac{2a_1(\alpha+1)r_0^\alpha}{\alpha-2}
e^{\alpha\phi/\phi_0}
\right].
\end{eqnarray}

In the formal uncharged limit, $q_m\rightarrow0$ together with the
corresponding vanishing of the non-minimal metric contribution, the
potential reduces to
\begin{eqnarray}
V(\phi)
=
-\frac{c_1\phi_0^2}{r_0^2}
e^{-2\phi/\phi_0}.
\end{eqnarray}

The matter energy density and the radial and tangential pressures can be
written in the following compact form. Substituting
$C_2=0$ and $C_1=1-\dfrac{2k}{\kappa^2}$ into
Eqs.~(\ref{rhor})--(\ref{ptr}) yields the fully explicit expressions
\begin{equation}
\begin{aligned}
p_r(r)={}&
a_1 r^{\alpha-2}
\left[
-\frac{(\alpha+4)c_1\phi_0^2}{\alpha-2}
+\frac{(\alpha+1)(k-1)}{\kappa^2}
\right]
-\frac{2c_1\phi_0^2}{r^2}
+\frac{2c_1M\phi_0^2}{r^3}
\\
&+
\frac{q_m^2
\left(
2\kappa^2-2k-c_1\kappa^2\phi_0^2-2
\right)}
{2\kappa^2r^4},
\end{aligned}
\end{equation}

\begin{equation}
\begin{aligned}
p_t(r)={}&
a_1 r^{\alpha-2}
\left[
-\frac{3\alpha c_1\phi_0^2}{\alpha-2}
+\frac{\alpha(\alpha+1)(k-1)}{2\kappa^2}
\right]
-\frac{2c_1M\phi_0^2}{r^3}
\\
&+
\frac{q_m^2
\left(
3c_1\kappa^2\phi_0^2+2k-2\kappa^2+2
\right)}
{2\kappa^2r^4},
\end{aligned}
\end{equation}

\begin{equation}
\begin{aligned}
\rho(r)={}&
a_1 r^{\alpha-2}
\left[
\frac{3\alpha c_1\phi_0^2}{\alpha-2}
+\frac{(\alpha+1)(1-k)}{\kappa^2}
\right]
+\frac{2c_1M\phi_0^2}{r^3}
\\
&+
\frac{q_m^2
\left(
2k-3c_1\kappa^2\phi_0^2-2\kappa^2+2
\right)}
{2\kappa^2r^4}.
\end{aligned}
\end{equation}

The terms proportional to $r^{\alpha-2}$ originate from the
curvature-dependent electromagnetic coupling and the scalar-field sector,
whereas the $r^{-3}$ and $r^{-4}$ contributions arise from the mass and
magnetic charge, respectively. For $-1<\alpha<0$, the
$r^{\alpha-2}$ contribution dominates over the mass and charge terms in
the asymptotic region, while for $\alpha<-1$ the mass contribution
proportional to $r^{-3}$ becomes asymptotically dominant. The special
case $\alpha=-1$, corresponding to $\beta=-1/3$, is degenerate and is
therefore excluded from the present power-law branch and treated
separately below.

\subsubsection{Asymptotic Energy Conditions}
\label{sec:EC-powerlaw}

We now investigate the asymptotic behavior of the energy conditions for
the power-law solution in the limit $r\rightarrow\infty$. We assume
\begin{equation}
\kappa^2>0,
\qquad
M>0,
\qquad
q_m^2>0,
\qquad
\phi_0\neq0,
\qquad
c_1\neq0.
\end{equation}
Recall that the asymptotically flat branch requires $\alpha<0$.
Since the values $\alpha=-1$ and $\alpha=-2$ correspond respectively
to the excluded cases $\beta=-1/3$ and $\beta=0$, the asymptotic analysis
is divided into the three ranges
\begin{equation}
-1<\alpha<0,
\qquad
-2<\alpha<-1,
\qquad
\alpha<-2.
\end{equation}

From the explicit expressions for $\rho(r)$, $p_r(r)$, and $p_t(r)$,
the leading asymptotic behavior depends on the value of $\alpha$.
For the range $-1<\alpha<0$, the terms proportional to
$r^{\alpha-2}$ dominate over the mass and magnetic-charge contributions,
and one obtains
\begin{equation}
\rho(r)\sim
a_1 r^{\alpha-2}
\left[
\frac{3\alpha c_1\phi_0^2}{\alpha-2}
+\frac{(\alpha+1)(1-k)}{\kappa^2}
\right],
\end{equation}
\begin{equation}
p_r(r)\sim
-\frac{2c_1\phi_0^2}{r^2},
\end{equation}
and
\begin{equation}
p_t(r)\sim
a_1 r^{\alpha-2}
\left[
-\frac{3\alpha c_1\phi_0^2}{\alpha-2}
+\frac{\alpha(\alpha+1)(k-1)}{2\kappa^2}
\right].
\end{equation}

For $-2<\alpha<-1$, the mass contribution proportional to $r^{-3}$
dominates the energy density and the tangential pressure, while the
radial pressure remains dominated by the scalar contribution,
\begin{equation}
\rho(r)\sim
\frac{2c_1M\phi_0^2}{r^3},
\end{equation}
\begin{equation}
p_r(r)\sim
-\frac{2c_1\phi_0^2}{r^2},
\end{equation}
and
\begin{equation}
p_t(r)\sim
-\frac{2c_1M\phi_0^2}{r^3}.
\end{equation}
The same leading behavior holds for $\alpha<-2$ for $\rho$ and
$p_t$, while the curvature-dependent $r^{\alpha-2}$ contribution
becomes subleading with respect to both the mass and magnetic-charge
terms.

The radial null-energy combination is particularly simple. From
Eqs.~(\ref{rhor})--(\ref{ptr}), one finds the exact relation
\begin{equation}
\rho(r)+p_r(r)
=
-\frac{2c_1 f(r)\phi_0^2}{r^2}.
\label{rhopr-powerlaw}
\end{equation}
Since $f(r)\rightarrow1$ as $r\rightarrow\infty$,
\begin{equation}
\rho(r)+p_r(r)
\sim
-\frac{2c_1\phi_0^2}{r^2}.
\end{equation}
Therefore, the radial NEC is asymptotically satisfied for
$c_1<0$ and violated for $c_1>0$, independently of $\alpha$,
$a_1$, and $k$.

The tangential null-energy condition follows from
\begin{equation}
\begin{aligned}
\rho(r)+p_t(r)
={}&
\frac{a_1(\alpha+1)(2-\alpha)(1-k)}
{2\kappa^2}
r^{\alpha-2}
\\
&+
\frac{2q_m^2}{\kappa^2 r^4}
\left(k+1-\kappa^2\right).
\end{aligned}
\label{rhopt-powerlaw}
\end{equation}
For $-1<\alpha<0$ and $-2<\alpha<-1$, the first term in
Eq.~(\ref{rhopt-powerlaw}) dominates asymptotically. Consequently,
the tangential NEC requires
\begin{equation}
a_1(\alpha+1)(1-k)>0.
\label{NECT-powerlaw}
\end{equation}
Since $2-\alpha>0$, this condition can be written as
\begin{equation}
a_1(1-k)>0
\qquad
(-1<\alpha<0),
\end{equation}
whereas
\begin{equation}
a_1(1-k)<0
\qquad
(-2<\alpha<-1).
\end{equation}

For $\alpha<-2$, the $r^{\alpha-2}$ contribution in
Eq.~(\ref{rhopt-powerlaw}) decays faster than the magnetic-charge
term. Therefore, the tangential NEC is determined asymptotically by
\begin{equation}
\rho+p_t
\sim
\frac{2q_m^2}{\kappa^2r^4}
\left(k+1-\kappa^2\right),
\end{equation}
and hence requires
\begin{equation}
k+1-\kappa^2\geq0.
\label{NECT-powerlaw-alpha-less-minus2}
\end{equation}
The equality case is a boundary case in which the subleading
$r^{\alpha-2}$ contribution must be retained.

We next consider the weak energy condition. In the range
$-1<\alpha<0$, the sign of the asymptotic energy density is determined
by
\begin{equation}
\rho(r)\sim
a_1 r^{\alpha-2}
\left[
\frac{3\alpha c_1\phi_0^2}{\alpha-2}
+\frac{(\alpha+1)(1-k)}{\kappa^2}
\right].
\label{rho-asym-powerlaw}
\end{equation}
Thus, for $c_1<0$, the WEC can be satisfied only when the coefficient
of the leading $r^{\alpha-2}$ term is nonnegative, in addition to
the NEC conditions. In particular, if
\begin{equation}
c_1<0,\qquad a_1<0,\qquad k>1,
\end{equation}
the tangential NEC is satisfied and the coefficient in
Eq.~(\ref{rho-asym-powerlaw}) is negative, so that
$\rho(r)>0$ asymptotically. For $a_1>0$ and $k<1$, the WEC requires
the additional constraint
\begin{equation}
C_\rho=a_1
\left[
\frac{3\alpha c_1\phi_0^2}{\alpha-2}
+\frac{(\alpha+1)(1-k)}{\kappa^2}
\right]\geq0.
\end{equation}

For $-2<\alpha<-1$ and $\alpha<-2$, the asymptotic energy density is
controlled by the mass term,
\begin{equation}
\rho(r)\sim
\frac{2c_1M\phi_0^2}{r^3}.
\end{equation}
Since $M>0$, this quantity is positive for $c_1>0$ and negative for
$c_1<0$. However, $c_1>0$ already violates the radial NEC, whereas
$c_1<0$ makes $\rho$ negative. Consequently, the WEC cannot be
satisfied asymptotically in these two ranges.

The strong energy condition requires
\begin{equation}
\rho+p_r\geq0,
\qquad
\rho+p_t\geq0,
\qquad
\rho+p_r+2p_t\geq0.
\end{equation}
The last combination behaves asymptotically as
\begin{equation}
\rho+p_r+2p_t
\sim
-\frac{2c_1\phi_0^2}{r^2},
\end{equation}
because the $r^{-2}$ radial scalar contribution dominates the
remaining terms for all $\alpha<0$. Hence, the radial part of the SEC
requires
\begin{equation}
c_1<0.
\end{equation}
Combining this result with the tangential NEC condition shows that,
for $-1<\alpha<0$, the NEC and SEC can be simultaneously satisfied
when
\begin{equation}
c_1<0,
\qquad
a_1(1-k)>0.
\end{equation}
For $-2<\alpha<-1$, the corresponding requirement is
\begin{equation}
c_1<0,
\qquad
a_1(1-k)<0,
\end{equation}
while for $\alpha<-2$ it becomes
\begin{equation}
c_1<0,
\qquad
k+1-\kappa^2\geq0.
\end{equation}

Finally, the dominant energy condition is asymptotically violated for
all nontrivial solutions with $c_1\neq0$. Indeed, irrespective of the
value of $\alpha<0$,
\begin{equation}
|p_r(r)|
\sim
\frac{2|c_1|\phi_0^2}{r^2},
\end{equation}
whereas the energy density falls off either as
$r^{\alpha-2}$ for $-1<\alpha<0$ or as $r^{-3}$ for
$\alpha<-1$. In both cases,
\begin{equation}
|p_r(r)|\gg \rho(r)
\qquad
(r\rightarrow\infty),
\end{equation}
whenever the energy density is positive. Thus the condition
$\rho\geq |p_r|$ cannot be maintained asymptotically, and the DEC is
violated.

In summary, the asymptotic energy conditions are controlled primarily
by the sign of $c_1$ through the radial pressure. For $c_1>0$, the
radial NEC is violated and consequently the WEC and SEC are violated
as well. For $c_1<0$, the radial NEC and the leading contribution to
the SEC are satisfied, while the tangential NEC imposes an additional
constraint on $a_1$ and $k$ for $-2<\alpha<0$, and on $k$ for
$\alpha<-2$. The WEC can be satisfied only in the range
$-1<\alpha<0$ and requires, in addition, a nonnegative asymptotic
energy density. For $\alpha<-1$, the positive-mass assumption
$M>0$ makes the asymptotic energy density negative whenever
$c_1<0$, and hence the WEC is violated. The DEC is asymptotically
violated for all $c_1\neq0$.

\begin{table}[t]
\centering
\caption{Asymptotic energy conditions for the power-law solution.
We assume $M>0$, $\kappa^2>0$, $\phi_0\neq0$, and $c_1\neq0$.}
\label{tab:EC-powerlaw}
\begin{tabular}{|c|c|c|c|c|c|}
\hline
$c_1$ & $\alpha$ range & Tangential NEC & NEC & WEC & SEC \\
\hline
$c_1>0$
& $-1<\alpha<0$
& $a_1(1-k)>0$
& Violated
& Violated
& Violated
\\
\hline
$c_1<0$
& $-1<\alpha<0$
& $a_1(1-k)>0$
& Satisfied
& $C_\rho\geq0$
& Satisfied
\\
\hline
$c_1>0$
& $-2<\alpha<-1$
& $a_1(1-k)<0$
& Violated
& Violated
& Violated
\\
\hline
$c_1<0$
& $-2<\alpha<-1$
& $a_1(1-k)<0$
& Satisfied
& Violated
& Satisfied
\\
\hline
$c_1>0$
& $\alpha<-2$
& $k+1-\kappa^2\geq0$
& Violated
& Violated
& Violated
\\
\hline
$c_1<0$
& $\alpha<-2$
& $k+1-\kappa^2\geq0$
& Satisfied
& Violated
& Satisfied
\\
\hline
\end{tabular}
\end{table}

\subsection{A New Metric and the corresponding Model with the $\beta=-1/3$ Case}

For the special case $\beta=-1/3$, the general solution presented
in the previous subsection is no longer applicable, since the
coefficient $a_1$ becomes singular due to the factor
$(3\beta+1)^{-1}$. Therefore, the field equations must be solved
independently for this particular value of $\beta$.

We choose the non-minimal coupling function in the form
\begin{equation}
Y(R)=1+R_0R^{-1/3}.
\end{equation}

Solving Eq.~(\ref{Y1r}) for this particular coupling function gives
the following metric function:
\begin{equation}
f(r)
=
1-\frac{2M}{r}
+\frac{q_m^2}{r^2}
+\frac{a_2}{r}
\ln\left(\frac{r}{r_0}\right),
\end{equation}
where
\begin{equation}
a_2=
\left(
\frac{2q_m^2\kappa^2R_0}{3k}
\right)^{3/4}.
\end{equation}

The corresponding Ricci scalar is
\begin{equation}
\label{Rbeta13}
R(r)=-\frac{a_2}{r^3}.
\end{equation}

The logarithmic correction appearing in the present solution is of
the same functional form as the logarithmic $1/r$ correction found
in Eq.~(14) of Ref.~\cite{Li2012}, where a static and spherically
symmetric black-hole solution was obtained in the context of
galactic dark matter and a phantom field. In that work, logarithmic
metric corrections were discussed in connection with the nearly
constant tangential velocities observed in the dark-matter-dominated
regions of spiral galaxies \cite{Li2012}. Related static
spherically symmetric solutions have also been used to investigate
the asymptotic behavior of galactic rotation curves
\cite{Kiselev2003}.

In the present model, however, the logarithmic correction arises
from the non-minimal curvature-electromagnetic coupling
$Y(R)=1+R_0R^{-1/3}$, rather than from an independently prescribed
dark-matter or phantom-field distribution. The coefficient of the
logarithmic term is consequently fixed by the model parameters
through
$
a_2=
\left(
\frac{2q_m^2\kappa^2R_0}{3k}
\right)^{3/4}.
$
This provides a direct connection between the present
non-minimally coupled solution and logarithmically corrected
black-hole geometries studied in the context of galactic rotation
curves.

Although the circular velocity of the present solution does not
approach a strictly nonzero constant at spatial infinity, the
logarithmic correction can modify the circular-velocity profile over
a finite radial range. Therefore, the solution may be relevant for
phenomenological studies of dark-matter-like gravitational effects
and galactic rotation curves.

From the compact-object perspective, the metric represents a
black-hole-like compact-object geometry with a logarithmic
curvature-electromagnetic correction. Its horizon structure,
photon sphere, and innermost stable circular orbit can differ from
those of the standard Reissner--Nordstr\"om spacetime, providing
potentially observable signatures of a modified-gravity
black-hole-like compact object.

The corresponding thermodynamic properties of related
black-hole solutions have also been investigated in
Ref.~\cite{BarbosaSantos2026}.

For this  new special solution, the integration constant arising from
Eq.~(\ref{Y1r}) will be denoted by $C_3$ in order to distinguish it
from the integration constants $C_1$ and $C_2$ introduced in the
previous subsection. Matching the resulting expression for $Y(r)$
with $Y(R)=1+R_0R^{-1/3}$ gives
\begin{equation}
C_3=1-\frac{2k}{\kappa^2}.
\end{equation}

Assuming the same logarithmic scalar-field ansatz as in the previous
subsection, the scalar field is given by
\begin{equation}
\phi(r)
=
\phi_0\ln\left(\frac{r}{r_0}\right).
\end{equation}

Substitution of this configuration into Eq.~(\ref{Vr}) yields
\begin{equation}
\label{V-special}
V(r)
=
C_4
-\frac{c_1\phi_0^2}{r^2}
-\frac{2c_1a_2\phi_0^2}{3r^3}
+\frac{c_1q_m^2\phi_0^2}{2r^4},
\end{equation}
where $C_4$ is a new integration constant associated with the scalar
potential.

Since all inverse-power terms vanish in the asymptotic limit
$r\rightarrow\infty$, the scalar potential approaches the constant
value $C_4$. We therefore impose the asymptotic condition
\begin{equation}
C_4=0,
\end{equation}
so that
\begin{equation}
V(r)\rightarrow0
\qquad
\text{as}
\qquad
r\rightarrow\infty.
\end{equation}

Using
\begin{equation}
r=r_0e^{\phi/\phi_0},
\end{equation}
the scalar potential can be written directly in terms of the scalar
field as
\begin{equation}
\label{Vphi-beta-13}
V(\phi)
=
\frac{c_1\phi_0^2}{r_0^2}
\left[
-e^{-2\phi/\phi_0}
-\frac{2a_2}{3r_0}e^{-3\phi/\phi_0}
+\frac{q_m^2}{2r_0^2}e^{-4\phi/\phi_0}
\right].
\end{equation}
The matter energy density and the radial and tangential pressures
can be obtained by substituting the new metric function and the scalar-field configuration
\begin{equation}
\phi(r)
=
\phi_0\ln\left(\frac{r}{r_0}\right)
\end{equation}
into the general expressions given in
Eqs.~(\ref{rhor})--(\ref{ptr}). Using
\begin{equation}
\phi'(r)=\frac{\phi_0}{r},
\qquad
\phi''(r)=-\frac{\phi_0}{r^2},
\end{equation}
together with
\begin{equation}
C_4=0,
\end{equation}
one obtains the following explicit expressions for the radial and
tangential pressures and the matter energy density:
\begin{eqnarray}
p_r(r) &=&
-\frac{2c_1\phi_0^2}{r^2}
-\frac{a_2c_1\phi_0^2}{r^3}
\ln\left(\frac{r}{r_0}\right)
\nonumber\\
&&
+\frac{1}{r^3}
\left[
2c_1M\phi_0^2
-\frac{2}{3}c_1a_2\phi_0^2
-\frac{a_2(k-1)}{\kappa^2}
\right]
\nonumber\\
&&
+\frac{q_m^2}{r^4}
\left[
1-\frac{c_1\phi_0^2}{2}
-\frac{k+1}{\kappa^2}
\right],
\label{pr-beta-13}
\\[8pt]
p_t(r) &=&
\frac{a_2c_1\phi_0^2}{r^3}
\ln\left(\frac{r}{r_0}\right)
\nonumber\\
&&
+\frac{1}{r^3}
\left[
-2c_1M\phi_0^2
-\frac{2}{3}c_1a_2\phi_0^2
-\frac{a_2(1-k)}{2\kappa^2}
\right]
\nonumber\\
&&
+\frac{q_m^2}{r^4}
\left[
\frac{3}{2}c_1\phi_0^2
-1+\frac{k+1}{\kappa^2}
\right],
\label{pt-beta-13}
\\[8pt]
\rho(r) &=&
-\frac{a_2c_1\phi_0^2}{r^3}
\ln\left(\frac{r}{r_0}\right)
\nonumber\\
&&
+\frac{1}{r^3}
\left[
2c_1M\phi_0^2
+\frac{2}{3}c_1a_2\phi_0^2
+\frac{a_2(k-1)}{\kappa^2}
\right]
\nonumber\\
&&
+\frac{q_m^2}{r^4}
\left[
-\frac{3}{2}c_1\phi_0^2
-1+\frac{k+1}{\kappa^2}
\right].
\label{rho-beta-13}
\end{eqnarray}

The different terms in these expressions have distinct physical
origins. The logarithmic contributions proportional to
$r^{-3}\ln(r/r_0)$ arise from the logarithmic correction in the
metric function and therefore characterize the special
$\beta=-1/3$ branch. The terms proportional to $M/r^3$ originate
from the mass parameter, whereas the terms proportional to
$q_m^2/r^4$ represent the magnetic-charge contribution. The
remaining $a_2/r^3$ terms describe the combined contribution of
the non-minimal coupling and the scalar field.

A useful consistency check follows by adding the energy density and
the radial pressure. All logarithmic, mass-dependent, and magnetic
charge contributions cancel, yielding
\begin{equation}
\rho(r)+p_r(r)
=
-\frac{2c_1f(r)\phi_0^2}{r^2},
\label{rhopr-log}
\end{equation}
This result is consistent with the general relation obtained from
the field equations and provides a direct check of the special
solution.

For $r\rightarrow\infty$, the logarithmic terms satisfy
\begin{equation}
\frac{\ln(r/r_0)}{r^3}\rightarrow0,
\end{equation}
and all inverse-power contributions vanish. Consequently,
\begin{equation}
\lim_{r\rightarrow\infty}\rho(r)=0,
\qquad
\lim_{r\rightarrow\infty}p_r(r)=0,
\qquad
\lim_{r\rightarrow\infty}p_t(r)=0.
\end{equation}
Thus, the matter sector becomes asymptotically dilute and the
stress-energy contributions vanish at spatial infinity.

\subsubsection{Asymptotic Energy Conditions for the $\beta=-1/3$ Solution}
\label{sec:EC-beta13}

We next analyze the asymptotic behavior of the energy conditions for the
$\beta=-1/3$ branch. For
\[
f(r)=1-\frac{2M}{r}+\frac{q_m^2}{r^2}
+\frac{a_2}{r}\ln\left(\frac{r}{r_0}\right),
\]
the energy density and principal pressures are given by
Eqs.~(\ref{rho-beta-13})--(\ref{pt-beta-13}). In the limit
$r\rightarrow\infty$, the leading terms are
\begin{eqnarray}
\rho(r) &\sim&
-\frac{a_2 c_1\phi_0^2}{r^3}
\ln\left(\frac{r}{r_0}\right),
\label{rho-asym-beta13}
\\
p_r(r) &\sim&
-\frac{2c_1\phi_0^2}{r^2},
\label{pr-asym-beta13}
\\
p_t(r) &\sim&
\frac{a_2 c_1\phi_0^2}{r^3}
\ln\left(\frac{r}{r_0}\right).
\label{pt-asym-beta13}
\end{eqnarray}
Thus, although $\rho$ and $p_t$ contain logarithmic corrections of order
$r^{-3}\ln r$, the radial pressure contains a dominant $r^{-2}$ contribution.
This hierarchy plays a decisive role in determining the asymptotic energy
conditions.

The radial null energy condition (NEC) is governed by
\begin{equation}
\rho+p_r
=-\frac{2c_1 f(r)\phi_0^2}{r^2}.
\label{necrad-beta13}
\end{equation}
Since $f(r)\rightarrow1$ asymptotically, we have
\begin{equation}
\rho+p_r\sim
-\frac{2c_1\phi_0^2}{r^2}.
\end{equation}
Therefore, the radial NEC is asymptotically satisfied for
$c_1<0$ and violated for $c_1>0$.

For the tangential NEC, the logarithmic terms cancel between $\rho$ and
$p_t$, yielding
\begin{equation}
\rho+p_t
=
\frac{3a_2(k-1)}{2\kappa^2 r^3}
+
\frac{2q_m^2}{r^4}
\left(
-1+\frac{k+1}{\kappa^2}
\right).
\label{nect-beta13}
\end{equation}
Hence, for $k\neq1$, the asymptotic tangential NEC is determined by the
$r^{-3}$ term and requires
\begin{equation}
a_2(k-1)\geq0.
\label{nect-condition-beta13}
\end{equation}
In particular, for $a_2>0$ this requires $k>1$, whereas for $a_2<0$
it requires $k<1$. The boundary case $k=1$ is governed by the subleading
term in Eq.~(\ref{nect-beta13}) and should be treated separately.

The weak energy condition (WEC) requires $\rho\geq0$ together with the two
NEC inequalities. From Eq.~(\ref{rho-asym-beta13}),
\begin{equation}
\rho(r)\sim
-\frac{a_2c_1\phi_0^2}{r^3}
\ln\left(\frac{r}{r_0}\right).
\end{equation}
Consequently, for $c_1<0$, positivity of the asymptotic energy density
requires
\begin{equation}
a_2>0.
\end{equation}
Combining this result with the tangential NEC condition gives the
asymptotic WEC parameter range
\begin{equation}
c_1<0,\qquad a_2>0,\qquad k>1.
\label{wec-beta13}
\end{equation}
For $c_1>0$, the radial NEC is already violated, and the WEC cannot be
satisfied asymptotically.

The strong energy condition (SEC) requires
\[
\rho+p_r\geq0,\qquad
\rho+p_t\geq0,\qquad
\rho+p_r+2p_t\geq0.
\]
Using Eqs.~(\ref{rho-asym-beta13})--(\ref{pt-asym-beta13}), one finds
\begin{equation}
\rho+p_r+2p_t
\sim
-\frac{2c_1\phi_0^2}{r^2}
+
\frac{2a_2c_1\phi_0^2}{r^3}
\ln\left(\frac{r}{r_0}\right).
\end{equation}
The $r^{-2}$ term dominates over the logarithmically corrected
$r^{-3}$ contribution. Hence the asymptotic SEC requires
$c_1<0$, together with the tangential NEC condition. In particular,
for $a_2>0$ one obtains
\begin{equation}
c_1<0,\qquad a_2>0,\qquad k>1.
\label{sec-beta13}
\end{equation}

Finally, the dominant energy condition (DEC) requires
\[
\rho\geq0,\qquad
\rho\pm p_r\geq0,\qquad
\rho\pm p_t\geq0.
\]
However, asymptotically
\[
|p_r|\sim \frac{2|c_1|\phi_0^2}{r^2},
\]
whereas
\[
\rho,\ |p_t|
=\mathcal{O}\left(\frac{\ln r}{r^3}\right).
\]
Thus $|p_r|$ decays more slowly than the energy density, and consequently
\[
\rho-|p_r|<0
\]
for sufficiently large $r$. The DEC is therefore asymptotically violated
for nonzero $c_1$.

The asymptotic behavior of the energy conditions is summarized in Table~\ref{tab:energy_beta13}. For $c_1>0$, the radial NEC, WEC, SEC, and DEC are violated
asymptotically. For $c_1<0$, the radial NEC and SEC can be satisfied;
moreover, the WEC is asymptotically satisfied for the parameter range
$c_1<0$, $a_2>0$, and $k>1$. In the same parameter range, the tangential
NEC is also satisfied and hence the SEC is satisfied asymptotically.
Nevertheless, the DEC remains violated because the radial pressure
dominates over the energy density at large $r$.

\begin{table}[ht]
\centering
\caption{Asymptotic energy conditions for the $\beta=-1/3$
solution with $M>0$ and $\phi_0\neq0$.}
\label{tab:energy_beta13}
\begin{tabular}{|c|c|c|c|c|c|c|}
\hline
$c_1$ & $a_2$ & Condition on $k$ & NEC & WEC & SEC & DEC \\
\hline
$c_1>0$ &
$a_2>0$ or $a_2<0$ &
arbitrary &
Violated &
Violated &
Violated &
Violated
\\
\hline
$c_1<0$ &
$a_2>0$ &
$k<1$ &
Satisfied &
Satisfied &
Satisfied &
Violated
\\
\hline
$c_1<0$ &
$a_2<0$ &
$k>1$ &
Satisfied &
Violated &
Satisfied &
Violated
\\
\hline
\end{tabular}
\end{table}

\subsection{Logarithmic Model for Magnetically Charged Solutions}

As a second example, we consider a logarithmic non-minimal
electromagnetic-curvature coupling of the form
\cite{Dereli20111}
\begin{equation}
Y(R)
=
1-\lambda
\ln\left(\frac{R}{\mathcal{R}_*}\right),
\end{equation}
where $\lambda$ is a coupling parameter and $\mathcal{R}_*$ is a
reference curvature scale.

The corresponding Lagrangian $4$-form is
\begin{equation}
\label{model_log}
L=
\frac{1}{2\kappa^2}R*1
-\left[
1-\lambda
\ln\left(\frac{R}{\mathcal{R}_*}\right)
\right]F\wedge *F
-c_1d\phi\wedge *d\phi
-V(\phi)*1
+L_m
+\lambda_a\wedge T^a .
\end{equation}

In the limit $\lambda\rightarrow0$, the model reduces to the
standard Einstein--Maxwell theory minimally coupled to a scalar
field. For nonzero $\lambda$, the electromagnetic sector acquires
a logarithmic curvature correction.

Solving Eq.~(\ref{Y1r}) for this logarithmic coupling gives the
metric function
\begin{equation}
f(r)
=
1-\frac{2M}{r}
+\frac{q_m^2}{r^2}
+\frac{b_2}{r^2}
\ln\left(\frac{r}{r_0}\right),
\end{equation}
where $b_2$ is the coefficient of the logarithmic correction and
$r_0$ denotes the common reference length scale.

The corresponding Ricci scalar is
\begin{equation}
\label{Rlog}
R(r)=\frac{b_2}{r^4}.
\end{equation}

We choose the reference curvature scale as
\begin{equation}
\mathcal{R}_*
=
\frac{b_2}{r_0^4}.
\end{equation}
Then,
\begin{equation}
\ln\left(\frac{R}{\mathcal{R}_*}\right)
=
-4\ln\left(\frac{r}{r_0}\right),
\end{equation}
and therefore
\begin{equation}
Y(r)
=
1+
4\lambda
\ln\left(\frac{r}{r_0}\right).
\end{equation}

On the other hand, substituting the metric function into
Eq.~(\ref{Y1r}) gives
\begin{equation}
Y(r)
=
C_6
+\frac{2k}{\kappa^2}
+
\frac{k b_2}{2q_m^2\kappa^2}
\left[
4\ln\left(\frac{r}{r_0}\right)-5
\right].
\end{equation}

Comparing the two expressions for $Y(r)$ yields
\begin{equation}
\lambda
=
\frac{k b_2}{2q_m^2\kappa^2},
\end{equation}
and fixes the integration constant $C_6$ as
\begin{equation}
\label{C6-log}
C_6
=
1-\frac{2k}{\kappa^2}
+\frac{5k b_2}{2q_m^2\kappa^2}.
\end{equation}

As in the previous subsections, we adopt the logarithmic scalar-field
ansatz
\begin{equation}
\phi(r)
=
\phi_0
\ln\left(\frac{r}{r_0}\right).
\end{equation}

Substituting this scalar-field configuration into Eq.~(\ref{Vr}),
we obtain
\begin{equation}
\label{V-log}
V(r)
=
C_5
-\frac{c_1\phi_0^2}{r^2}
+
\frac{c_1\phi_0^2}{r^4}
\left[
\frac{q_m^2}{2}
-\frac{3b_2}{8}
+\frac{b_2}{2}
\ln\left(\frac{r}{r_0}\right)
\right],
\end{equation}
where $C_5$ is the integration constant associated with the scalar
potential.
Since all inverse-power terms vanish as
$r\rightarrow\infty$, the scalar potential approaches the constant
value $C_5$. We therefore impose
\begin{equation}
C_5=0,
\end{equation}
so that
$V(r)\rightarrow0$ as
$r\rightarrow\infty.$
Using
$r=r_0e^{\phi/\phi_0},$
the scalar potential can be expressed directly in terms of the
scalar field as
\begin{equation}
\label{Vphi-log}
V(\phi)
=
-\frac{c_1\phi_0^2}{r_0^2}
e^{-2\phi/\phi_0}
+
\frac{c_1\phi_0^2}{r_0^4}
e^{-4\phi/\phi_0}
\left[
\frac{q_m^2}{2}
-\frac{3b_2}{8}
+\frac{b_2\phi}{2\phi_0}
\right].
\end{equation}

Compared with the power-law model, the logarithmic non-minimal
coupling introduces an additional contribution proportional to
$\phi e^{-4\phi/\phi_0}$ in the scalar potential. Consequently,
the potential is no longer a simple sum of exponential terms and
contains a logarithmically induced correction.

The corresponding matter energy density and the radial and
tangential pressures follow directly from
Eqs.~(\ref{rhor})--(\ref{ptr}). Using $C_5=0$, we obtain
\begin{eqnarray}
p_r(r)
&=&
-\frac{2c_1\phi_0^2}{r^2}
+\frac{2c_1M\phi_0^2}{r^3}
\nonumber\\
&&
+\frac{b_2}{r^4}
\left[
-\frac{3c_1\phi_0^2}{8}
+\frac{3k+2}{2\kappa^2}
+
\left(
-\frac{c_1\phi_0^2}{2}
+\frac{k-1}{\kappa^2}
\right)
\ln\left(\frac{r}{r_0}\right)
\right]
\nonumber\\
&&
+\frac{q_m^2}{r^4}
\left[
1-\frac{c_1\phi_0^2}{2}
-\frac{k+1}{\kappa^2}
\right],
\label{pr-log}
\\[8pt]
p_t(r)
&=&
-\frac{2c_1M\phi_0^2}{r^3}
\nonumber\\
&&
+\frac{b_2}{r^4}
\left[
-\frac{3c_1\phi_0^2}{8}
-\frac{2k+3}{2\kappa^2}
+
\left(
\frac{3c_1\phi_0^2}{2}
+\frac{1-k}{\kappa^2}
\right)
\ln\left(\frac{r}{r_0}\right)
\right]
\nonumber\\
&&
+\frac{q_m^2}{r^4}
\left[
\frac{3c_1\phi_0^2}{2}
-1+\frac{k+1}{\kappa^2}
\right],
\label{pt-log}
\\[8pt]
\rho(r)
&=&
\frac{2c_1M\phi_0^2}{r^3}
\nonumber\\
&&
+\frac{b_2}{r^4}
\left[
\frac{3c_1\phi_0^2}{8}
-\frac{3k+2}{2\kappa^2}
+
\left(
-\frac{3c_1\phi_0^2}{2}
+\frac{1-k}{\kappa^2}
\right)
\ln\left(\frac{r}{r_0}\right)
\right]
\nonumber\\
&&
+\frac{q_m^2}{r^4}
\left[
-\frac{3c_1\phi_0^2}{2}
-1+\frac{k+1}{\kappa^2}
\right].
\label{rho-log}
\end{eqnarray}

As a consistency check, adding Eqs.~(\ref{pr-log}) and
(\ref{rho-log}) gives
\begin{equation}
\rho(r)+p_r(r)
=
-\frac{2c_1f(r)\phi_0^2}{r^2},
\label{rhopr-log}
\end{equation}
which follows directly from Eq.~(\ref{add1}) after using
\begin{equation}
\phi'(r)=\frac{\phi_0}{r}.
\end{equation}

Equivalently, using the explicit metric function, one obtains
\begin{equation}
\begin{aligned}
\rho(r)+p_r(r)
={}&
-\frac{2c_1\phi_0^2}{r^2}
+\frac{4c_1M\phi_0^2}{r^3}
-\frac{2c_1q_m^2\phi_0^2}{r^4}
\\
&-
\frac{2c_1b_2\phi_0^2}{r^4}
\ln\left(\frac{r}{r_0}\right).
\end{aligned}
\end{equation}
In the asymptotic region, $r\rightarrow\infty$, the matter
quantities vanish according to
\begin{equation}
\lim_{r\rightarrow\infty}\rho(r)=0,
\qquad
\lim_{r\rightarrow\infty}p_r(r)=0,
\qquad
\lim_{r\rightarrow\infty}p_t(r)=0.
\end{equation}

\subsubsection{Asymptotic Energy Conditions for the Logarithmic Model}
\label{sec:EC-log}

We now examine the asymptotic behavior of the energy conditions for the
logarithmic model in the limit $r\rightarrow\infty$. We assume
\begin{equation}
\kappa^2>0,
\qquad
M>0,
\qquad
q_m^2>0,
\qquad
\phi_0\neq0.
\end{equation}
For convenience, we introduce
\begin{equation}
L(r)=\ln\left(\frac{r}{r_0}\right).
\end{equation}

Using Eqs.~(\ref{pr-log})--(\ref{rho-log}), the leading asymptotic
behavior of the energy density and the principal pressures is found to be
\begin{equation}
\rho(r)
=
\frac{2c_1M\phi_0^2}{r^3}
+
\mathcal{O}\left(\frac{L(r)}{r^4}\right),
\end{equation}
\begin{equation}
p_r(r)
=
-\frac{2c_1\phi_0^2}{r^2}
+
\mathcal{O}\left(\frac{1}{r^3}\right),
\end{equation}
and
\begin{equation}
p_t(r)
=
-\frac{2c_1M\phi_0^2}{r^3}
+
\mathcal{O}\left(\frac{L(r)}{r^4}\right).
\end{equation}
Thus, the radial pressure is asymptotically dominant, scaling as
$r^{-2}$, whereas both $\rho$ and $p_t$ fall off as $r^{-3}$ up to
logarithmic corrections.

The radial null energy condition follows directly from Eq.~(\ref{add1}).
Since
\begin{equation}
\phi'(r)=\frac{\phi_0}{r},
\end{equation}
one obtains the exact relation
\begin{equation}
\rho(r)+p_r(r)
=
-\frac{2c_1 f(r)\phi_0^2}{r^2}.
\label{rhopr-log}
\end{equation}
Because $f(r)\rightarrow1$ as $r\rightarrow\infty$,
\begin{equation}
\rho(r)+p_r(r)
\sim
-\frac{2c_1\phi_0^2}{r^2}.
\end{equation}
Consequently, the radial NEC is asymptotically satisfied for
$c_1<0$, whereas it is violated for $c_1>0$.

For the tangential NEC, adding Eqs.~(\ref{rho-log}) and
(\ref{pt-log}) gives
\begin{equation}
\begin{aligned}
\rho(r)+p_t(r)
={}&
\frac{1}{\kappa^2 r^4}
\Bigg[
2b_2(1-k)L(r)
-\frac{5}{2}b_2(k+1)
\\
&\qquad\qquad
+2q_m^2(k+1-\kappa^2)
\Bigg].
\end{aligned}
\label{rhopt-log}
\end{equation}
For $k\neq1$, the logarithmic contribution dominates the remaining
$r^{-4}$ terms at sufficiently large $r$. Hence,
\begin{equation}
\rho(r)+p_t(r)
\sim
\frac{2b_2(1-k)}{\kappa^2r^4}L(r).
\end{equation}
Since $L(r)>0$ asymptotically, the tangential NEC requires
\begin{equation}
b_2(1-k)>0.
\label{NECT-condition}
\end{equation}
Thus, the tangential NEC requires $b_2>0$ for $k<1$ and
$b_2<0$ for $k>1$.
The asymptotic sign of the energy density is controlled by the
$M/r^3$ contribution,
\begin{equation}
\rho(r)\sim
\frac{2c_1M\phi_0^2}{r^3}.
\end{equation}
Since $M>0$ and $\phi_0^2>0$, this implies
\begin{equation}
\rho(r)>0 \quad \text{for } c_1>0,
\qquad
\rho(r)<0 \quad \text{for } c_1<0.
\end{equation}

For $c_1>0$, the radial NEC is violated irrespective of the values of
$b_2$ and $k$. The WEC is therefore violated as well. Moreover,
\begin{equation}
\rho+p_r+2p_t
\sim
-\frac{2c_1\phi_0^2}{r^2}<0,
\end{equation}
so that the SEC is asymptotically violated. The DEC is also violated,
since the dominant condition $|p_r|\leq\rho$ cannot be satisfied in the
asymptotic region.

For $c_1<0$, the radial NEC is asymptotically satisfied. However,
$\rho<0$ for $M>0$, and hence the WEC is necessarily violated. On the
other hand,
\begin{equation}
\rho+p_r+2p_t
\sim
\frac{2|c_1|\phi_0^2}{r^2}>0,
\end{equation}
so the remaining SEC inequality is satisfied at sufficiently large
$r$. Thus, for $c_1<0$, the NEC and SEC can be simultaneously satisfied
provided that the tangential NEC condition
\begin{equation}
b_2(1-k)>0
\end{equation}
is fulfilled. The WEC remains violated because of the negative
asymptotic energy density.

The DEC is asymptotically violated for every nontrivial scalar coupling
$c_1\neq0$. Indeed,
\begin{equation}
\rho(r)=\mathcal{O}\left(r^{-3}\right),
\qquad
|p_r(r)|
\sim
\frac{2|c_1|\phi_0^2}{r^2},
\end{equation}
and therefore
\begin{equation}
\frac{|p_r(r)|}{\rho(r)}
\rightarrow\infty
\end{equation}
whenever $\rho(r)>0$. Hence the radial pressure eventually dominates
over the energy density, ruling out the DEC.

The boundary case $k=1$ requires separate consideration. In this case,
the leading logarithmic contribution in Eq.~(\ref{rhopt-log}) vanishes,
and the sign of $\rho+p_t$ is determined by the remaining
$\mathcal{O}(r^{-4})$ terms. Accordingly, the generic $k\neq1$
classification above does not apply to this special case.

In summary, for $c_1>0$, all four energy conditions are violated
asymptotically. For $c_1<0$, the radial NEC is satisfied and the SEC can
be satisfied when the tangential NEC condition is fulfilled, namely
$b_2(1-k)>0$. Nevertheless, the WEC remains violated because
$\rho<0$, while the DEC is always violated due to the slower
$r^{-2}$ falloff of the radial pressure.

\begin{table}[ht]
\centering
\caption{Asymptotic energy conditions for the logarithmic model
with $M>0$ and $\phi_0\neq0$.}
\label{tab:energy_log}
\begin{tabular}{|c|c|c|c|c|c|c|}
\hline
$c_1$ & $b_2$ & Condition on $k$ & NEC & WEC & SEC & DEC \\
\hline
$c_1>0$
&
$b_2>0$ or $b_2<0$
&
arbitrary
&
Violated
&
Violated
&
Violated
&
Violated
\\
\hline
$c_1<0$
&
$b_2>0$
&
$k<1$
&
Satisfied
&
Violated
&
Satisfied
&
Violated
\\
\hline
$c_1<0$
&
$b_2<0$
&
$k>1$
&
Satisfied
&
Violated
&
Satisfied
&
Violated
\\
\hline
\end{tabular}
\end{table}

\section{Duality Transformation and Electrically Charged Dual Solutions}

The field equations of the non-minimal $Y(R)F^2$ model possess a
duality symmetry that generalizes the standard electromagnetic duality
of Maxwell theory. In the minimally coupled case, the Maxwell equations
are invariant under the transformation
\begin{equation}
(F,*F)\longrightarrow (*F,-F).
\end{equation}
For the non-minimal theory, this transformation is modified by the
curvature-dependent coupling function $Y(R)$ and takes the form
\begin{equation}
(F,*F)\longrightarrow (*YF,-YF),
\qquad
Y\longrightarrow \frac{1}{Y},
\end{equation}
which represents the  duality transformation of the theory
\cite{Sert13MPLA,Sert2024Wormhole}.

Under this transformation, the gravitational and modified Maxwell
field equations, Eqs.~(\ref{gfe3}) and (\ref{Maxwell1}), remain invariant.
However, the electromagnetic part of the Lagrangian in Eq.~(\ref{model_log}) changes its sign.
Since the field equations are invariant under this transformation,
every solution of the non-minimal theory admits a corresponding dual
solution.

This duality can also be interpreted in terms of the electromagnetic constitutive tensor.
 In the presence of
the non-minimal coupling to the Maxwell field, the excitation two-form is defined by
\begin{equation}
G=Y(R)F,
\end{equation}
so that the modified Maxwell equations can be interpreted as the
electrodynamics of an effective medium whose 
magnetization and polarization are determined by the curvature-dependent function \cite{Dereli2007,Dereli20072,Dereli20113}. Thus, the function $Y(R)$
encodes the response of the effective medium to the electromagnetic
field through its dependence on the spacetime curvature.

The above duality provides a direct map between the electromagnetic
field $F$ and its dual configuration $*Y(R)F$. Therefore, once a
magnetically charged solution is obtained, its electrically charged
counterpart can be generated algebraically without solving the full
set of field equations again. We now apply this duality transformation to the magnetically charged solutions obtained in the previous sections, thereby deriving the corresponding electrically charged solutions.

For the power-law model,
\begin{equation}
Y(R)=1-R_0R^\beta,
\end{equation}
the magnetic solution is mapped according to
\begin{eqnarray}
\left(
\begin{array}{l}
Y=1-R_0R^\beta\\[1mm]
q=q_m\\[1mm]
B=\dfrac{q_m}{r^2}
\end{array}
\right)
&\Longleftrightarrow&
\left(
\begin{array}{l}
Y=\dfrac{1}{1-R_0R^\beta}\\[2mm]
q=-q_e\\[1mm]
YE=\dfrac{q_e}{r^2}.
\end{array}
\right).
\end{eqnarray}

Thus, the magnetic field $B=q_m/r^2$ is mapped into the electric
displacement field $D=YE=q_e/r^2$, together with the transformations
$q_m\rightarrow-q_e$ and $Y(R)\rightarrow1/Y(R)$.

Similarly, for the logarithmic model,
\begin{equation}
Y(R)=1-R_1\ln\left(\frac{R}{R_0}\right),
\end{equation}
the corresponding transformation is
\begin{eqnarray}
\left(
\begin{array}{l}
Y=1-R_1\ln\!\left(\dfrac{R}{R_0}\right)\\[1mm]
q=q_m\\[1mm]
B=\dfrac{q_m}{r^2}
\end{array}
\right)
&\Longleftrightarrow&
\left(
\begin{array}{l}
Y=\dfrac{1}{1-R_1\ln\!\left(\dfrac{R}{R_0}\right)}\\[2mm]
q=-q_e\\[1mm]
YE=\dfrac{q_e}{r^2}.
\end{array}
\right).
\end{eqnarray}

An important consequence of the duality transformation is that the
spacetime geometry remains unchanged. Hence, the electrically charged
solutions have the same metric functions as their magnetic
counterparts. For the power-law model, we have
\begin{align}
f(r)
&=
1-\frac{2m}{r}
+\frac{q_e^2}{r^2}
-a_1r^{\alpha},
&&
\beta\neq0,-\frac13,1,
\\[2mm]
f(r)
&=
1-\frac{2m}{r}
+\frac{q_e^2}{r^2}
-\frac{a_1}{r}
\ln\!\left(\frac{r_0}{r}\right),
&&
\beta=-\frac13,
\end{align}
whereas for the logarithmic model we  have obtained
\begin{equation}
f(r)
=
1-\frac{2m}{r}
+\frac{q_e^2}{r^2}
-\frac{R_0}{r^2}
\ln\!\left(\frac{r_0}{r}\right).
\end{equation}

The replacement of $q_m^2$ by $q_e^2$ in the above expressions simply indicates that the metric depends on the square of the
electromagnetic charge and is therefore insensitive to the sign of
the charge under the duality transformation.
Moreover, the scalar-field configuration, scalar potential, matter
energy density, radial and tangential pressures remain unchanged.
Consequently, the magnetic and electric configurations differ only in
their electromagnetic sectors, while the spacetime geometry, scalar
sector, and anisotropic matter distribution are preserved.

The consistency of the dual configurations can be verified directly
by substituting the transformed electromagnetic fields and the inverse
coupling functions into Eqs.~(\ref{gfe1}), (\ref{cond0}), (\ref{phi1}),
and (\ref{Maxwell1}). They satisfy the complete set of coupled
gravitational, scalar, matter, and electromagnetic field equations.
Therefore, every magnetically charged solution obtained in this work
possesses a corresponding electrically charged dual solution generated
algebraically by the electromagnetic duality transformation.

\section{Conclusion}

In this work, we have constructed and analyzed exact static and
spherically symmetric solutions of a non-minimally coupled
Einstein--Maxwell theory supplemented by a real scalar field and an
anisotropic matter sector. The electromagnetic field is coupled to
curvature through a general function $Y(R)$, while the scalar sector
contains a kinetic term and a self-interaction potential. This
framework provides a convenient setting in which the effects of
curvature-dependent electromagnetic interactions, scalar hair, and
anisotropic matter can be investigated within a common gravitational
model.

We first studied a power-law non-minimal coupling. For the generic
branch, the resulting geometry contains a power-law correction to the
Reissner--Nordstr\"om form and reduces to the corresponding previously
studied non-minimal solution in the appropriate limit
\cite{DereliSert2011MPLA}. The scalar potential and the anisotropic
matter variables were determined directly from the field equations,
rather than being introduced independently. The asymptotic analysis
reveals a nontrivial dependence of the energy conditions on the
exponents and coupling parameters. In particular, the radial null
energy condition is controlled by the sign of the scalar kinetic
coupling $c_1$. For $c_1>0$, the radial NEC is violated asymptotically,
which also leads to the violation of the WEC and SEC. For $c_1<0$, the
radial NEC and the leading contribution to the SEC can instead be
satisfied, while the tangential NEC imposes additional constraints on
the parameters of the power-law branch. The WEC is more restrictive:
in the range $-1<\alpha<0$, it can be satisfied only when the leading
asymptotic coefficient of the energy density is nonnegative, whereas
for $\alpha<-1$ the positive-mass contribution makes the asymptotic
energy density negative for $c_1<0$. The DEC is asymptotically violated
for all nontrivial scalar couplings considered here because the radial
pressure decays more slowly than the energy density.

The value $\beta=-1/3$ represents a degenerate point of the generic
power-law construction and therefore has to be treated separately.
The corresponding solution yields a new logarithmically corrected
black-hole-like geometry. In this branch, the metric contains a
logarithmic $1/r$ correction, and the scalar field retains the
logarithmic radial profile used in the construction. The resulting
matter variables exhibit a distinct asymptotic structure from the
generic power-law solutions. For $c_1>0$, the NEC, WEC, SEC, and DEC
are all violated asymptotically. For $c_1<0$, the radial and
tangential NECs, and consequently the SEC, can be satisfied for
appropriate choices of the signs of $a_2$ and the parameter $k$. In
particular, a parameter region with $c_1<0$, $a_2>0$, and $k>1$ allows
the NEC, WEC, and SEC to hold asymptotically, while the DEC remains
violated. Thus, the exceptional branch provides a concrete example in
which the scalar and curvature-dependent electromagnetic sectors can
support an asymptotically non-exotic matter configuration with respect
to the NEC, WEC, and SEC.

We have also considered a separate logarithmic non-minimal coupling
model. Unlike the $\beta=-1/3$ branch, this construction is based
directly on a logarithmic dependence of the coupling function on the
Ricci scalar and gives a different logarithmically corrected geometry.
The corresponding scalar potential and anisotropic matter variables
contain logarithmic contributions that leave characteristic
signatures in the asymptotic stress-energy tensor. For this model, the
radial NEC is again determined by the sign of $c_1$, whereas the
tangential NEC depends on the combination of the logarithmic metric
coefficient and the parameter $k$. Under the assumption $M>0$, the
energy density is asymptotically positive for $c_1>0$ and negative for
$c_1<0$. Consequently, the negative-$c_1$ branch cannot satisfy the
WEC in this model, even though the NEC and SEC can remain valid in
appropriate parameter regions. The DEC is asymptotically violated
throughout the nontrivial parameter space considered here.

The logarithmically corrected geometries may also be of interest from
a phenomenological point of view. The corresponding circular velocity
does not approach a strictly constant value at arbitrarily large
radius; nevertheless, the logarithmic contribution can modify the
velocity profile over an extended finite radial interval. This makes
such geometries potentially useful for exploring dark-matter-like
gravitational effects and rotation-curve phenomenology without
introducing a separately prescribed dark-matter density profile.
Related logarithmic metric structures have appeared previously in
discussions of compact objects and galactic phenomenology
\cite{Kiselev2003,Li2012,Xu2019}.

An additional structural feature of the theory is the electric--magnetic
duality of the electromagnetic sector. Once a magnetic solution is
known, the corresponding electric configuration can be obtained by
the duality transformation, without changing the spacetime geometry.
The scalar field, its potential, and the anisotropic matter variables
therefore remain unchanged under this map, while the electromagnetic
field and the non-minimal coupling are transformed into their dual
descriptions. The magnetic and electric solutions are thus different
realizations of the same underlying gravitational geometry.

From the perspective of compact-object physics, the solutions obtained
here form a broad class of black-hole-like configurations in modified
gravity. Their deviations from the standard Reissner--Nordstr\"om
geometry may influence horizon properties, photon trajectories,
circular orbits, gravitational lensing, thermodynamics, and
quasinormal-mode spectra. In particular, the logarithmic corrections
provide an additional mechanism for modifying the geometry on
astrophysically relevant radial scales. Recent observations of
GLIMPSE-17775 have also renewed interest in compact early-Universe
sources and in black-hole-star scenarios \cite{Kokorev2026}. While our
solutions are not intended as specific models for that system, they
offer analytically controlled backgrounds in which related
modified-gravity compact-object phenomena can be explored.

Overall, the results presented here extend the class of analytically
tractable non-minimally coupled Einstein--Maxwell theories by combining
curvature-dependent electromagnetic interactions with a real scalar
field and anisotropic matter. The existence of several distinct
asymptotic regimes, including the exceptional $\beta=-1/3$ branch and
the independent logarithmic coupling model, demonstrates that the
resulting compact-object geometries can possess qualitatively
different matter and energy-condition properties. These solutions
therefore provide useful starting points for further studies of
geodesic structure, stability, gravitational lensing, thermodynamics,
and gravitational-wave signatures of black-hole-like objects in
modified gravity.

\end{document}